\documentclass[12pt]{article}
\usepackage{epsfig}
\usepackage{axodraw}
 
\newcommand{\as}{\alpha_{\mathrm{s}}}
\newcommand{\aem}{\alpha_{\mathrm{em}}}
\newcommand{\stw}{\sin^2 \! \theta_W}
\renewcommand{\d}{\mathrm{d}}
\newcommand{\e}{\mathrm{e}}
\newcommand{\g}{\mathrm{g}}
\newcommand{\p}{\mathrm{p}}
\newcommand{\q}{\mathrm{q}}
\newcommand{\W}{\mathrm{W}}
\newcommand{\Z}{\mathrm{Z}}
\newcommand{\qbar}{\mathrm{\overline{q}}}
\newcommand{\pbar}{\mathrm{\overline{p}}}
\newcommand{\kT}{k_{\perp}}
\newcommand{\pT}{p_{\perp}}
\newcommand{\pTW}{p_{\perp\mathrm{W}}}
\newcommand{\mW}{m_{\mathrm{W}}}
\newcommand{\shat}{\hat{s}}
\newcommand{\that}{\hat{t}}
\newcommand{\uhat}{\hat{u}}

\newcommand{\lessim}{\raisebox{-0.8mm}%
{\hspace{1mm}$\stackrel{<}{\sim}$\hspace{1mm}}}

{\end{list}}
\newcounter{enumct}
\newenvironment{Enumerate}{\begin{list}{\arabic{enumct}.}%
{\usecounter{enumct}\setlength{\topsep}{0.2mm}%
\setlength{\partopsep}{0.2mm}\setlength{\itemsep}{0.2mm}%
\setlength{\parsep}{0.2mm}}}{\end{list}}
 
\newlength{\abstwidth}
\newlength{\captivewidth}
\newcommand{\captive}[1]{\rule{5mm}{0mm}%
\begin{minipage}{\captivewidth}%
\caption[small]{#1}\end{minipage}}
 
\begin{document}
 
\sloppy
 
\pagestyle{empty}
 
\begin{flushright}
LU TP 98--30 \\
December 1998
\end{flushright}
 
\vspace{\fill}
 
\begin{center}
{\LARGE\bf W Production in an Improved}\\[3mm]
{\LARGE\bf Parton-Shower Approach}\\[10mm]
{\Large Gabriela Miu\footnote{gabriela@thep.lu.se}
and Torbj\"orn Sj\"ostrand\footnote{torbjorn@thep.lu.se}}\\ [2mm]
{\it Department of Theoretical Physics,}\\[1mm]
{\it Lund University, Lund, Sweden}
\end{center}
 
\vspace{\fill}
\begin{center}
{\bf Abstract}\\[2ex]
\begin{minipage}{\abstwidth}
In the description of the production properties of gauge 
bosons ($\W^{\pm}$, $\Z^0$, $\gamma^*$) at colliders,
the lowest-order graph normally is not sufficient. 
The contributions of higher orders can be introduced either 
by an explicit order-by-order matrix-element calculation,
by a resummation procedure or by a parton-shower algorithm.
Each approach has its advantages and disadvantages.
We here introduce a method that allows the parton-shower
algorithm to be augmented by higher-order information,
thereby offering an economical route to a description of
all event properties. It is tested by comparing with 
the $\pT$ spectrum of $\W$ bosons at the Tevatron.
\end{minipage}
\end{center}

\vspace{\fill}
 
\clearpage
\pagestyle{plain}
\setcounter{page}{1}

The $\W^{\pm}$ and $\Z^0$ bosons have been extensively studied at 
colliders, in order to test the standard model \cite{PDG}. In 
recent years they have also made their debut as backgrounds to other 
processes of interest: top studies, Higgs searches, and so on.
Here it is  often the association of the $\W/\Z$ with one or 
several jets that is the source of concern. Such higher-order
corrections to the basic processes also serve as tests of QCD. 
It is therefore of some interest to improve the accuracy with 
which gauge boson production can be described. 

In this letter we will take the $\W^{\pm}$ production at hadron
colliders as a test bed to develop some ideas in this direction.
Specifically, we will discuss how to improve the lowest-order
description $\q\qbar' \to \W^{\pm}$ by a merging of the
first-order matrix elements $\q\qbar' \to \g\W^{\pm}$
and $\q\g \to \q'\W^{\pm}$ with a leading-log parton shower.
However, the formalism is valid for all colourless massive vector 
gauge bosons within and beyond the standard model: $\gamma^*$,
$\Z^0$, $\Z'^0$, $\W'^{\pm}$, and so on. It also applies e.g.
in $\e^+\e^- \to \gamma\Z^0$. One could in addition imagine 
extensions to quite different processes, such as Higgs production 
by $\g\g \to \mathrm{h}^0$, but this would require further study. 

The outline of the letter is the following. First we discuss
various approaches to $\W$ production, and their respective 
limitations. Then we zoom in on the shower method and
introduce a matrix-element-motivated method to improve it.
Finally we compare with data, specifically the $\W$ 
transverse-momentum spectrum at the Tevatron, and draw some 
conclusions.

In essence, one may distinguish three alternative descriptions 
of $\W$ production:
\begin{Enumerate}
\item \textit{Order-by-order matrix elements.} By a systematic 
expansion in powers of $\as$, a quite powerful machinery 
is obtained. For instance, there are calculations of the total 
Drell-Yan cross section to second order \cite{secondordersigma} and 
for the associated production of a $\W$ and up to four partons at 
the Born level \cite{Wplusfourjets}. A main problem is that there is 
no smooth transition between the different event classes, as one 
parton becomes soft or two partons collinear. The method is therefore 
better suited for exclusive questions than for an inclusive view of 
$\W$ event properties. 
\item \textit{Resummed matrix elements.} Here the effects of 
multiple parton emission are resummed, in impact-parameter or 
transverse-momentum space \cite{resummed}. Inclusive quantities 
such as the $\pTW$ spectrum can be well described in such an 
approach, but it should be noted that a nonperturbative input 
is required. The standard formalism does not give the exclusive 
set of partons accompanying the $\W$, however, and internally
does not respect correct kinematics.
\item \textit{Parton showers.} The parton-shower approach 
generates complete events, with correct kinematics. An arbitrary
number of partons is obtained, with a smooth and physical transition
between event classes ensured by the use of Sudakov form factors. 
On the other hand, the shower approach is formally only to 
leading-log accuracy (although many detailed choices are made to 
maximize agreement with next-to-leading-log results), and the 
description of the rate of exclusive parton configurations may 
be poor.  
\end{Enumerate}
Finally, note that the perturbative partonic stage is not observable
in experiment, but instead the hadronic jet one. Traditional
hadronization descriptions, such as string fragmentation 
\cite{AGIS}, are intended to be universal if
applied at some low cut-off scale $Q_0 \sim 1$~GeV of the
perturbative phase. This perfectly matches the shower
approach, but causes problems in the use of matrix elements. 

Given their complementary strengths, it is natural to attempt
a marriage of the matrix-element and parton-shower methods,
where the rate of well-separated jets is consistent with the
former while the substructure of jets is described by the latter.
The simpler solution, \textit{matching}, is to introduce a 
transition from one method to the other at some intermediate 
scale \cite{matching,mike,herwig}. Such an approach is convenient for 
descriptions of exclusive jet topologies, but tend to suffer
from discontinuities between event classes and around the
transition scale. More ambitious is the \textit{merging}
strategy, where matrix-element information is integrated into
the shower in such a way as to obtain a uniform and smooth
description. This approach so far has only been implemented
for the merging to $O(\as)$ of $\e^+\e^- \to \q\qbar$ with 
$\e^+\e^- \to \q\qbar\g$ \cite{merging,mike}. We will here introduce
a corresponding $O(\as)$ merging in hadronic $\W$ production.
Further details may be found in \cite{miu}.

Since we neglect the decay of the $\W$, alternatively imagine it
decaying leptonically, all QCD radiation occurs in the initial 
state. We will base our approach on the initial-state shower 
algorithm of \cite{isrshower}, as implemented in {\sc Pythia} 
\cite{pythia}. The principle of \textit{backwards evolution}
implies that a shower may be reconstructed by starting at the 
large $Q^2$ scale of the hard process and then gradually 
considering emissions at lower and lower virtualities, i.e.
earlier and earlier in the cascade chain (and in time).

The starting point is the standard DGLAP evolution equation 
\cite{DGLAP},
\begin{equation}
\frac{\d f_b(x,t)}{\d t} = \sum_a \int_x^1 \frac{\d x'}{x'} \, 
\frac{\as(t)}{2\pi} \, f_{a}(x',t) \, P_{a\to bc}(z) ~,
\end{equation}
with $f_i$ the distribution function of parton species $i$,
$x$ the momentum fraction carried by the parton, 
$t = \ln(Q^2/\Lambda^2_{\mathrm{QCD}})$ the resolution scale,
and $P_{a\to bc}(z)$ the AP splitting kernels for parton $b$
obtaining a fraction $z = x / x'$ of the $a$ momentum.
Normally the evolution is in terms of increasing $t$, but in
the backwards evolution $t$ is instead decreasing. Then 
the DGLAP equation expresses the rate at which partons $b$ of 
momentum fraction $x$ are `unresolved' into partons $a$ of 
fraction $x'$, in a step $\d t$ backwards. The corresponding 
relative probability is $\d P_b/\d t = (1/f_b) \, (\d f_b/\d t)$. 
The probability that $b$ remains resolved from some initial scale 
$t_{\mathrm{max}}$ down to $t < t_{\mathrm{max}}$ is thereby 
obtained by a Sudakov form factor
\begin{eqnarray}
S_b(x,t;t_{\mathrm{max}}) &=& \exp\left( -\int_t^{t_{\mathrm{max}}} 
\frac{1}{f_b(x,t')} \, \frac{\d f_b(x,t')}{\d t'} \, \d t' \right)
\nonumber \\
&=& \exp\left( -\int_t^{t_{\mathrm{max}}} \d t' \sum_a \int_x^1
\frac{\d x'}{x'} \, \frac{\as(t')}{2\pi} \, 
\frac{f_a(x',t')}{f_b(x,t')} \, P_{a\to bc}(z) \right) 
\nonumber \\
&=& \exp\left( -\int_t^{t_{\mathrm{max}}} \d t' \, 
\frac{\as(t')}{2\pi} \sum_a \int_x^1 \d z  \, 
\frac{x' f_a(x',t')}{x f_b(x,t')} \, P_{a\to bc}(z) \right) ~.
\label{sudakov}
\end{eqnarray}
From this expression it is a matter of standard Monte Carlo 
techniques to generate the complete branching $a \to bc$ 
\cite{isrshower}; e.g., the $t$ distribution of the branching is 
$-\d S_b(x,t;t_{\mathrm{max}}) / \d t$. Given parton $a$, one may 
in turn reconstruct which parton branched into it, and so on, 
down to the starting scale $Q_0$. In each branching, the $t$ scale 
gives the $t_{\mathrm{max}}$ value of the branching to be considered 
next, i.e. the $Q^2$ values are assumed strictly ordered.

\begin{figure}[tb]
\begin{center}
\begin{picture}(400,75)(0,0)
\ArrowLine(10,60)(110,60)
\LongArrow(110,60)(155,66)
\ArrowLine(110,60)(165,54)
\Text(140,47)[]{3}
\LongArrow(165,54)(190,64)
\Text(175,68)[]{4}
\ArrowLine(165,54)(200,38)
\Text(180,37)[]{1}
\ArrowLine(390,60)(300,60)
\LongArrow(300,60)(270,66)
\ArrowLine(300,60)(240,54)
\Text(270,47)[]{5}
\LongArrow(240,54)(220,64)
\Text(233,68)[]{6}
\ArrowLine(240,54)(200,38)
\Text(220,35)[]{2}
\LongArrow(200,38)(195,06)
\Text(215,15)[]{0(W)}
\end{picture}
\end{center}
\caption{Schematic picture of an initial-state parton shower,
extending from both sides of the event in to the $\W$. 
\label{ps-notation} }
\end{figure}
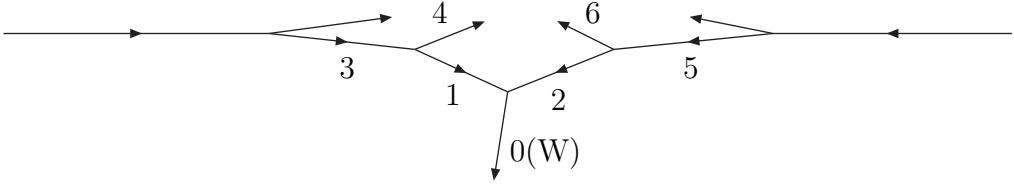

The definition of the $Q^2$ and $z$ variables is not unambiguous.
Referring to the notation of Fig.~\ref{ps-notation}, and to the
branching $3 \to 1 + 4$, the $Q^2$ scale in our algorithm 
\cite{isrshower} is associated with the 
spacelike virtuality of the produced parton 1, $Q^2 = - p_1^2$,
while $z$ is given by the reduction of squared invariant mass of
the contained subsystem, $z = (p_1 + p_2)^2/(p_3 + p_2)^2$.
In the limit of collinear kinematics, $Q^2=0$, one recovers the 
momentum fraction $z = p_1/p_3$. The $z$ definition couples the
two sides of the events, so that the order in which the
branchings $3 \to 1 + 4$ and $5 \to 2 + 6$ are considered makes
some difference for the final configuration. The rule adopted is 
therefore to reconstruct branching kinematics strictly in order 
of decreasing $Q^2$, i.e. interleaving emissions on the two sides 
of the event.

Now let us compare the step from $\q\qbar' \to \W$ to 
$\q\qbar' \to \g\W$ between the matrix-element and parton-shower
languages. Since only one branching is to be considered, 
the comparison has to be with a truncated shower,
e.g. where only the branching $3 \to 1 + 4$ occurs in 
Fig.~\ref{ps-notation}. The $2 \to 2$ process thus is
$\q(3) + \qbar'(2) \to \g(4) + \W(0)$, for which
\begin{eqnarray}
\shat &=& (p_3 + p_2)^2 = \frac{(p_1 + p_2)^2}{z} = \frac{\mW^2}{z} ~, 
\nonumber \\
\that &=& (p_3-p_4)^2 = p_1^2 = -Q^2 ~,
\label{stQz}
\\
\uhat &=& \mW^2 - \shat - \that = Q^2 - \frac{1-z}{z} \, \mW^2 ~.
\nonumber
\end{eqnarray}       

The matrix element for $\q\qbar' \to \g\W$ can be
written as \cite{Wsigma}
\begin{equation}
\left. \frac{\d \hat{\sigma}}{\d \that} \right|_{\mathrm{ME}} =
\frac{\sigma_0}{\shat} \, \frac{\as}{2\pi} \, \frac{4}{3} 
\, \frac{\that^2+\uhat^2+2\mW^2\shat}{\that\uhat} ~.
\label{MEgW}
\end{equation}
Here $\sigma_0$ is the cross section for $\q\qbar' \to \W$, 
$\sigma_0 = (\pi^2\aem/3\stw\mW^2) |V_{\q\qbar'}|^2 
\delta(1 - \mW^2 / x_1 x_2 s)$ in the narrow-width
limit, with $\delta(1 - \mW^2 / x_1 x_2 s) \mapsto
\int \d z \, \delta(1 - \mW^2 / z x_3 x_2 s)$
in the $2 \to 2$ process kinematics. (The 
details of the $\sigma_0$ factor are not relevant for the point 
we want to make, so the presentation is intentionally sketchy.)
Now rewrite eq.~(\ref{MEgW}) in terms of $z$ and $Q^2$, using 
eq.~(\ref{stQz}):
\begin{eqnarray}
\left. \frac{\d \hat{\sigma}}{\d Q^2} \right|_{\mathrm{ME}} & = &
\frac{\sigma_0 z}{\mW^2} \, \frac{\as}{2\pi} \, \frac{4}{3} 
\, \frac{(1+z^2)\mW^4 - 2z(1-z)Q^2\mW^2+2 z^2 Q^4}
{z Q^2 ((1-z)\mW^2 - z Q^2)} 
\nonumber \\
& \stackrel{Q^2 \to 0}{\longrightarrow} &
\sigma_{0} \, \frac{\as}{2\pi} \, \frac{4}{3} \, 
\frac{1+z^2}{1-z} \, \frac{1}{Q^2} =
\left. \frac{\d \hat{\sigma}}{\d Q^2} \right|_{\mathrm{PS1}} ~.
\end{eqnarray} 
We here easily recognize the splitting kernel for $\q \to \q\g$,
i.e. the matrix element reduces to the the normal shower
expression in the collinear limit, as it should be. Some
extra but trivial work is necessary to include the convolution with
parton distributions, which involves $f_1(x_1, Q^2)$ in lowest order 
and $f_3(x_3,Q^2)$ for the $O(\as)$ processes.

In order to study how the shower populates the phase space,
it is straightforward to translate back the above expression,
\begin{equation}
\left. \frac{\d \hat{\sigma}}{\d \that} \right|_{\mathrm{PS1}} =
\frac{\sigma_0}{\shat} \, \frac{\as}{2\pi} \, \frac{4}{3} 
\, \frac{\shat^2+ \mW^4}{\that(\that+\uhat)} ~.
\end{equation}
To this we should add the other possible shower history,
where the gluon is emitted by a branching $5 \to 2 + 6$ instead;
after all, the matrix-element expression contains both amplitudes.
The collinear singularity $Q^2 \to 0$ here corresponds to emission
along direction 2 rather than direction 1.
In that case the r\^oles of $\that$ and $\uhat$ are interchanged,
and the cross section $\d \hat{\sigma} / \d \that |_{\mathrm{PS2}}$ 
is easily obtained. The total shower rate is given by the sum,
\begin{equation}
\left. \frac{\d \hat{\sigma}}{\d \that} \right|_{\mathrm{PS}} =
\left. \frac{\d \hat{\sigma}}{\d \that} \right|_{\mathrm{PS1}} +
\left. \frac{\d \hat{\sigma}}{\d \that} \right|_{\mathrm{PS2}} =
\frac{\sigma_0}{\shat} \, \frac{\as}{2\pi} \, \frac{4}{3} 
\, \frac{\shat^2+ \mW^4}{\that\uhat} ~.
\end{equation}
Thus the singularity structure of the parton-shower and matrix-element
rates agree, giving a ratio
\begin{equation}
R_{\q\qbar' \to \g\W}(\shat,\that) = 
\frac{ (\d\hat{\sigma}/\d\that)_{\mathrm{ME}} }%
     { (\d\hat{\sigma}/\d\that)_{\mathrm{PS}} } = 
\frac{\that^2+\uhat^2+2 \mW^2\shat}{\shat^2+\mW^4} =
1 - \frac{2\that\uhat}{\shat^2+\mW^4}
\end{equation}
constrained to the range
\begin{equation}
\frac{1}{2} < R_{\q\qbar' \to \g\W}(\shat,\that) \le 1 ~.
\end{equation}

The same exercise may be carried out for $\q\g \to \q'\W$:
\begin{eqnarray}
\left. \frac{\d \hat{\sigma}}{\d \that} \right|_{\mathrm{ME}} & = &
\frac{\sigma_0}{\shat} \, \frac{\as}{2\pi} \, \frac{1}{2} 
\, \frac{\shat^2+\uhat^2+2\mW^2\that}{- \shat\uhat}
\label{MEqW}  
\nonumber \\
& = & \frac{\sigma_0 z}{\mW^2} \, \frac{\as}{2\pi} \, \frac{1}{2} 
\, \frac{(z^2 + (1-z)^2) \mW^4 + 2 z^2 Q^2 \mW^2 + z^2 Q^4}{z Q^2 \mW^2}
\nonumber \\
& \stackrel{Q^2 \to 0}{\longrightarrow} &
\sigma_0 \, \frac{\as}{2\pi} \, \frac{1}{2} \, 
(z^2 + (1-z)^2) \, \frac{1}{Q^2} =
\left. \frac{\d \hat{\sigma}}{\d Q^2} \right|_{\mathrm{PS}} ~,
\\
\left. \frac{\d \hat{\sigma}}{\d \that} \right|_{\mathrm{PS}} & = &
\frac{\sigma_0}{\shat} \, \frac{\as}{2\pi} \, \frac{1}{2} 
\, \frac{\shat^2+ 2 \mW^2 (\that + \uhat)}{-\shat\uhat} ~,
\\
R_{\q \g \to \q' \W}(\shat,\that) & = &
\frac{ (\d\hat{\sigma}/\d\that)_{\mathrm{ME}} }%
     { (\d\hat{\sigma}/\d\that)_{\mathrm{PS}} } = 
\frac{\shat^2 + \uhat^2 + 2 \mW^2 \that}{\shat^2 + 2 \mW^2 (\that + \uhat)} 
= 1 + \frac{\uhat(\uhat -2\mW^2)}{(\shat-\mW^2)^2 + \mW^4} ~,
\\
& & 1 \leq R_{\q\g \to \q' \W}(\shat,\that) \leq 
\frac{\sqrt{5}-1}{2(\sqrt{5}-2)} <  3.
\end{eqnarray}
Note that, unlike the $\q \qbar' \to \g \W$ process, there is no 
addition of two shower histories when comparing with matrix 
elements, since here also the latter contains two separate terms 
corresponding to $\q\g$ and $\g\q$ initial states, respectively.

The $\q \qbar' \to \g\W$ process receives contributions from
two Feynman graphs, $t$-channel and $u$-channel, and the shower
thus exactly matches this set, although obviously it does not 
include interference between the two. The $\q\g \to \q'\W$
process is different, since only its $u$-channel graph is covered
by the parton-shower formalism, while the $s$-channel one has no
correspondence. Since this latter graph is free from collinear 
singularities, the shower is not misbehaving in any regions of 
phase space because of this omission, but it is interesting to
speculate that the larger value for 
$R_{\q \g \to \q' \W}(\shat,\that)$ than for 
$R_{\q\qbar' \to \g\W}(\shat,\that)$ partly may have its origin here
(remember that a larger $R(\shat,\that)$ means a smaller shower 
emission rate).

Based on the above exercise, the standard parton-shower approach 
may be improved in two steps. The first is to note that, since the
shower so closely agrees with the correct matrix-element expression
--- much better than one might have had reason to expect ---
it is safe to apply the shower to all of phase space, i.e.
to have $Q_{\mathrm{max}}^2 \approx s$ rather than the more
traditional shower-generator limit $Q_{\mathrm{max}}^2 \approx \mW^2$
\cite{pythia,herwig}. The older choice was inspired in part
by the fear of a completely erroneous behaviour for $Q^2 \gg \mW^2$, 
in part by the typical factorization scale used for parton 
distributions in $\W$ cross-section formulae. Such a scale choice 
can be motivated by doublecounting arguments. Most easily this is 
seen in pure QCD processes, where a $2 \to 3$ process such as 
$\g\g \to \q\qbar\g$ could be obtained starting from several 
different $2 \to 2$ processes,
and classification by the hardest (most virtual) subgraph is
necessary to resolve ambiguities. Correspondingly, a $\W$
production graph could be reclassified once some parton has
$Q^2 > \mW^2$. That is, in general, one would have to consider 
QCD processes where the emission of a $\W$ is allowed as a 
`parton shower' correction. (The $s$-channel graph in $\q\g \to \q'\W$
is an example of this kind.) Doublecounting is not an issue, however, 
once we decide to represent the full $\W$ cross section by 
$\q\qbar \to \W$. (Remember that the shower does not change total
cross sections.)

The second step is to use standard Monte Carlo techniques to
correct branchings in the shower by the relevant ratio 
$R(\shat,\that)$, to bring the shower parton-emission rate in better 
agreement with the matrix-element one. This correction is applied 
to the branching closest to the hard scattering, i.e. with largest
virtuality, on both sides of the event, i.e. for $3 \to 1+ 4$ and 
$5 \to 2 + 6$ in Fig.~\ref{ps-notation}. By analogy with results for
time-like showers \cite{mike}, one could attempt to formulate 
more precise rules for when to apply corrections, but this one 
should come close enough and is technically the simplest solution.
(For instance, while our cascade is ordered in $Q^2$ rather than
in $\pT^2$, the emission with largest $\pT^2$ normally coincides 
with the largest $Q^2$ one, so either criterion for when to apply 
a correction would give very similar results.)
For a $\q \to \q\g$ shower branching, where the correction factor 
$R(\shat,\that) = R_{\q\qbar' \to \g\W}(\shat,\that) \le 1$, a 
candidate branching selected according to the Sudakov factor in 
eq.~(\ref{sudakov}) is then accepted with a probability 
$R(\shat,\that)$. In case of failure, the evolution
downwards in $Q^2$ is continued from the scale that failed
(the `veto algorithm', ensuring the correct form of the Sudakov).  
For a $\g\to\q\qbar$ branching, the fact that
$R(\shat,\that) = R_{\q\g \to \q'\W}(\shat,\that) \ge 1$ means 
that the procedure above cannot be used directly. Instead the 
normal $\g\to\q\qbar$ branching rate is enhanced by an ad hoc 
factor of 3, and the acceptance rate instead given by 
$R(\shat,\that)/3 < 1$.
  
Even with this injection of matrix-element information into the
parton shower, it is important to recognize that the shower still 
is different. The hardest emission is given by the matrix-element 
expression \textit{times} the related Sudakov form factor, thus
ensuring a smooth $\pT$ spectrum that vanishes in the limit 
$\pT \to 0$. By the continued shower history (without any
matrix-element corrections), further emissions
pick up where the Sudakov factor suppresses the hardest one, 
giving a total $\pT$ spectrum of emitted partons that is peaked
at the lower $\pT$ cut. This total spectrum is similar to the
matrix-element one, but deviates from it in that the shower 
includes kinematical and dynamical effects of gradually having
partons at larger and larger $x$ values and possibly of different
species at each softer emission. In some respects, it thus
provides a more sophisticated approach to resummation for the
properties of the recoiling $\W$. It also gives exclusive final 
states, including the possibility for the emitted partons (such as 
4 and 6 in Fig.~\ref{ps-notation}) to branch in their turn.

The parton shower redistributes the $\W$'s in phase space but does 
not change the total $\W$ cross section. It is thus feasible to  
use a higher-order calculation of this cross section as starting 
point, although we did not do it here. If higher orders enhance the 
total cross section by a factor $K$ (with $K$ a function e.g. of 
rapidity) relative to the lowest-order $\q\qbar' \to \W$ one, the 
implication of eqs.~(\ref{MEgW}) and (\ref{MEqW}) is that the one-jet 
rate is enhanced by the same factor. If instead another factor $K'$ 
is wanted here, the respective $R(\shat,\that)$ weight could then be 
modified by a factor $K'/K$. Note, however, that it is more difficult 
to introduce such a $K'/K$ factor consistently, since it presupposes 
a common definition between showers and matrix elements of what 
constitutes a jet.

\begin{figure}[tb]
\mbox{\epsfig{file=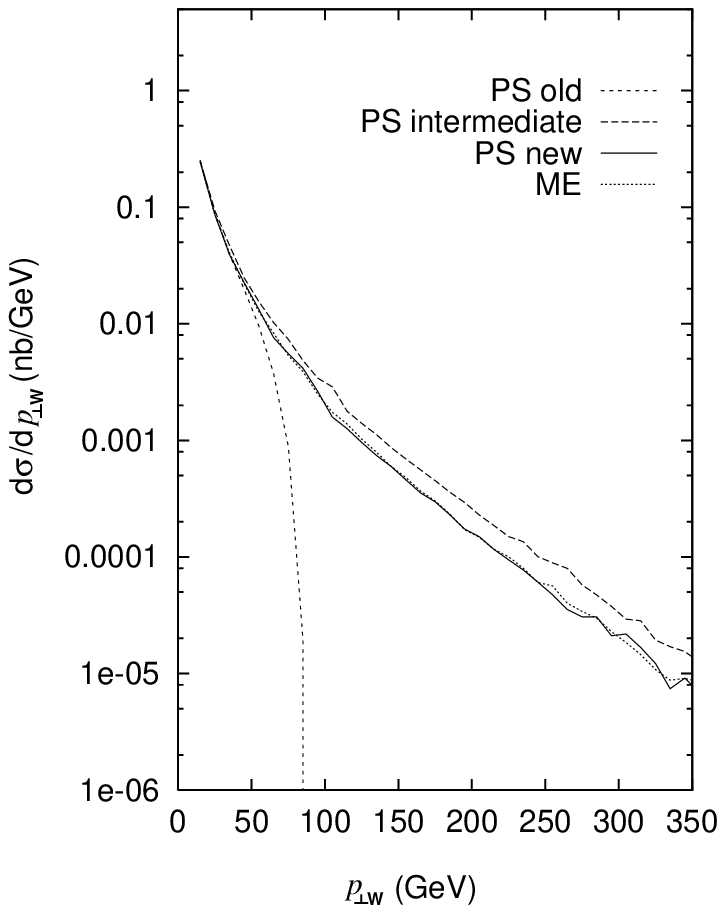,width=7.9cm}}
\mbox{\epsfig{file=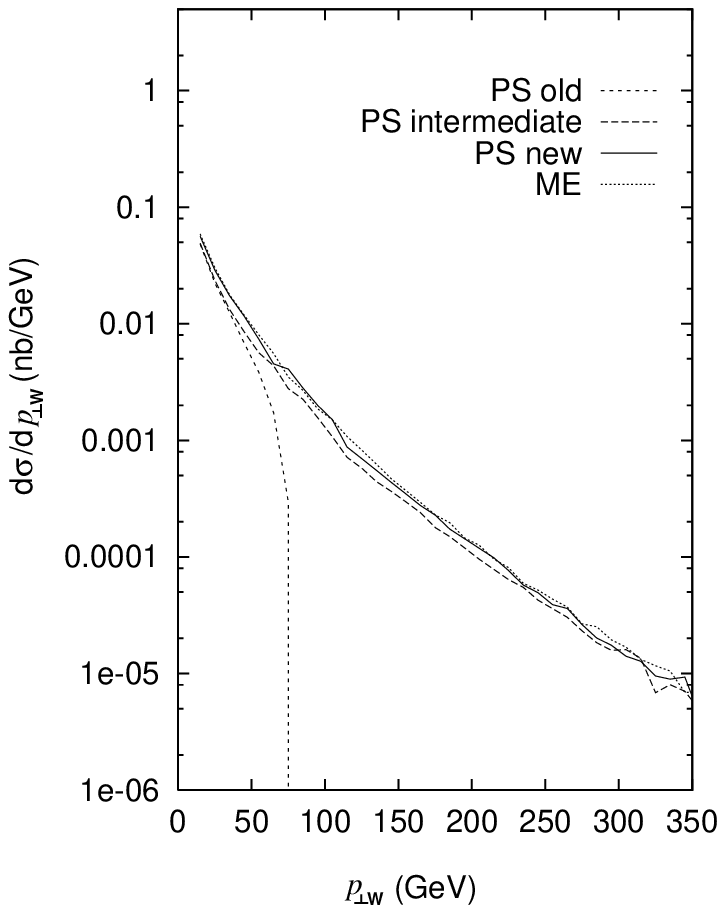,width=7.9cm}}
{ \mbox{ \begin{picture}(400,20)(0,0)\Text(123,2)[]{(a)}
\Text(352,2)[]{(b)} \end{picture} } }
\captive{The $\pTW$ distribution in $\p\pbar$ collisions at 1.8~TeV.
Parton distributions and $\as$ are frozen and only one emission 
at a time is allowed in the shower. Events are classified either as
(a) $\q\qbar' \to \g\W$ or (b) $\q\g \to \q'\W$.
\label{check} }
\end{figure}

As a first check, we want to confirm that the shower algorithm works 
as intended, reproducing the matrix-element expressions where it 
should. 
Thus the shower is artificially modified so as only to generate one 
branching at a time. In order to eliminate the influence of the 
Sudakov and the change of kinematics by previously considered 
emissions, the shower is restarted from each $Q^2$ actually selected 
above the cut, but returning to the original kinematics for 
$\q\qbar' \to \W$. Furthermore parton distributions and $\as$ are 
frozen, so as to avoid any scale-choice mismatches. The resulting $\W$ 
transverse momentum spectrum is shown in Fig.~\ref{check}, classified 
by the two possible branchings. In the old scheme, with 
$Q_{\mathrm{max}}^2 = \mW^2$, the drop of the $\pTW$ spectrum at
$\pTW \approx \mW$ is easily visible. Already the 
modification to $Q_{\mathrm{max}}^2 = s$ (the `intermediate' curves)
brings a marked improvement, and the further introduction of the 
$R(\shat,\that)$ weighting (`new') results in good agreement between 
the shower and the matrix elements.

\begin{figure}[tb]
\mbox{\epsfig{file=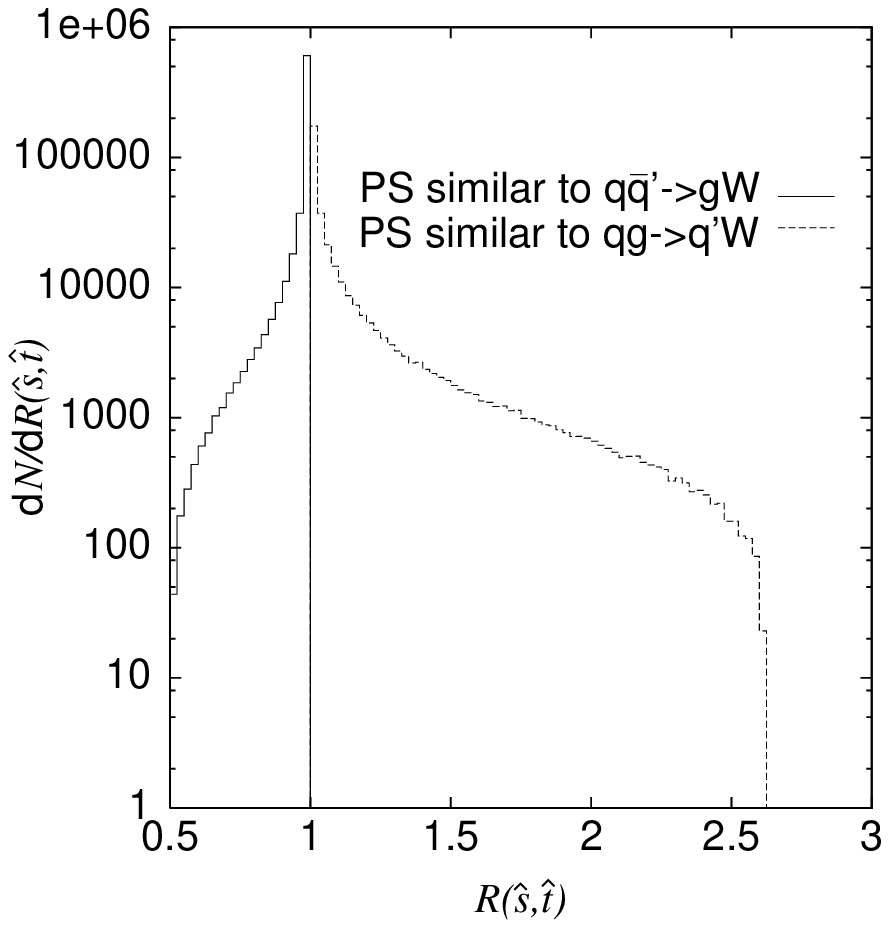,width=7.9cm}}
\mbox{\epsfig{file=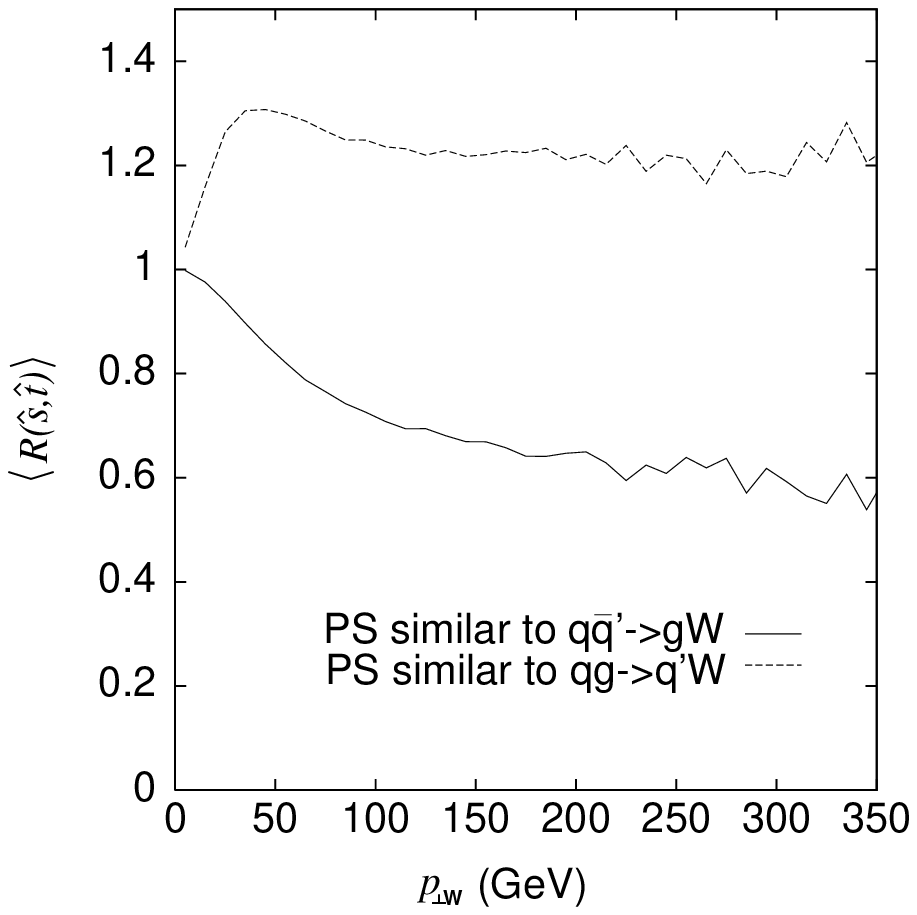,width=7.9cm}}
{ \mbox{ \begin{picture}(400,20)(0,0)\Text(123,2)[]{(a)}
\Text(352,2)[]{(b)} \end{picture} } }
\captive{$R(\shat,\that)$ distributions in $\p\pbar$ collisions at 
1.8~TeV. (a) The inclusive distribution. (b) The average value as 
a function of $\pTW$.  Events are classified either as
$\q\qbar' \to \g\W$ or $\q\g \to \q'\W$.
\label{Rfactor} }
\end{figure}

The $R(\shat,\that)$ factors are further studied in Fig.~\ref{Rfactor}.
It is seen that $R(\shat,\that)$ is close to unity for most of the
branchings (note the logarithmic scale). Also that 
$R(\shat,\that) \to 1$ as $\pT \to 0$, in accordance with the
demonstrated agreement of the parton shower and matrix elements in the
collinear limit. At large $\pT$ values, the $R(\shat,\that)$ factors
enhance the importance of the $\q\g \to \q'\W$ process relative
to the $\q\qbar' \to \g\W$ one by about a factor of 2. When the two
processes are not separated, the partial cancellation of having one 
$R(\shat,\that)$ a bit above unity and the other a bit below leads
to a rather modest net correction to $\pT$ spectra.

\begin{figure}[tb]
\begin{center}
\mbox{\epsfig{file=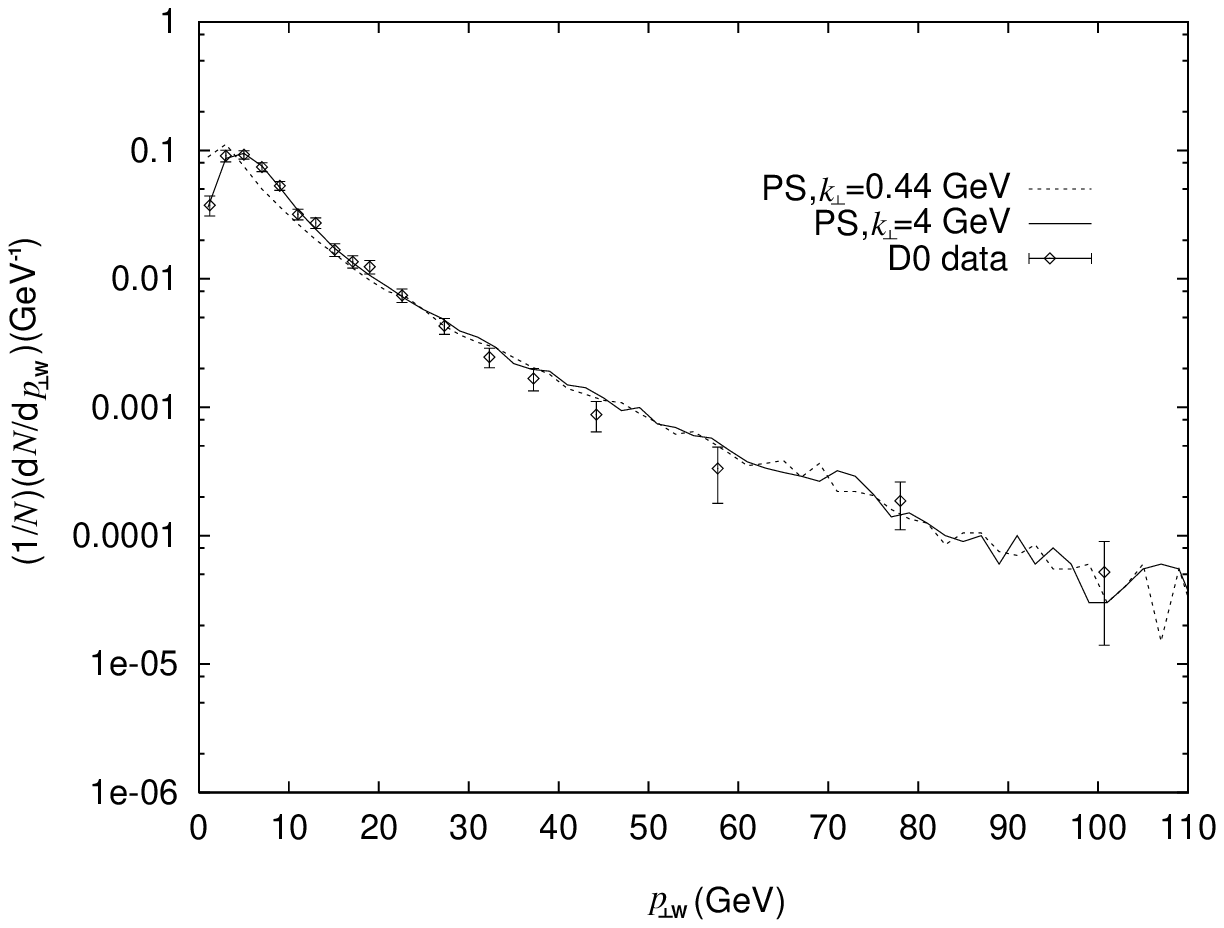,width=12cm}}
\end{center}
\captive{ 
Transverse momentum spectrum of the $\W$; full parton-shower
results compared with data from the D0 Collaboration \cite{D0data}.
(The error bars include both statistical and systematic uncertainties.)
\label{datacomp} }
\end{figure}
 
We now present results for the new parton shower, in all its complexity, 
i.e. with normal complete showers augmented by the two-step correction
process described above. In Fig.~\ref{datacomp} the shower $\pTW$
distribution is compared with experimental data from the D0 
collaboration \cite{D0data}. The agreement is good for large $\pTW$, 
but the shape at small $\pTW$ is rather different, with less activity
in the shower than in data. These results are for the default Gaussian 
primordial $\kT$ spectrum of width 0.44~GeV, as would be the order 
expected from a purely nonperturbative source related to confinement
inside the incoming hadrons. By now several indications have 
accumulated that a larger width is needed, however \cite{primkt},
although the origin of such an excess is not at all understood.
One hypothesis is that some radiation is overlooked by an imperfect 
modelling of the perturbative QCD radiation around or below the $Q_0$ 
cut-off scale. Whatever the reason, we may quantify the disagreement
by artificially increasing the primordial $\kT$ width. 
Fig.~\ref{datacomp} shows that an excellent agreement can be 
obtained, at all $\pTW$ values, with a 4~GeV width. The $\pTW$ 
distribution is essentially unchanged at large values, i.e. 
only the region $\pTW \lessim 20$~GeV is affected. In order to put 
the 4~GeV number in perspective, it should be noted that this is 
introduced as a `true' primordial $\kT$, i.e. carried by the parton
on each incoming hadron side that initiates the initial-state shower
at the $Q_0$ scale. If such a parton has an original momentum fraction
$x_0$ and the parton at the end of the cascade (that actually 
produces the $\W$) has fraction $x$, the $\W$ only receives a 
primordial $\kT$ kick scaled down by a factor $x/x_0$. In the 
current case, and also including the fact that two sides contribute, 
this translates into a rms width of 2.1~GeV for 
the primordial $\kT$ kick given to the $\W$. This number actually 
is not so dissimilar from values typically used in resummation 
descriptions \cite{resummed}.   

In summary, we see that it is possible to obtain a good description
of the complete $\pTW$ spectrum by fairly straightforward improvements
of a normal parton-shower approach. Corresponding improvements can
also be expected for the production of jets in association with the
$\W$. Especially, the good matching offered to hadronization descriptions
in this approach allows a complete simulation of the final state,
including the addition of a (possibly process-dependent) underlying
event. This way it should be possible to address e.g. the ratio of 
events with/without a jet accompanying the $\W$, where CDF and D0 have
obtained partly conflicting results, however using different jet 
definitions \cite{Wjet}. 

We end by reiterating that the formalism presented here is universal,
in the sense that the formulae in this paper are not unique for the
$\W$, but shared by all vector gauge bosons, after an appropriate 
replacement of $\mW$ and the constants in the $\sigma_0$ prefactor.   
Specifically, the reweighting factors $R(\shat,\that)$ need only
be modified to reflect the mass of the current resonance. The method
therefore should offer an accurate and economical route to the 
prediction of kinematical distributions for a host of new particles, 
to be searched for at the Tevatron and the LHC. It is also likely
that similar approaches can be developed for other classes of 
processes.


\begin{thebibliography}{99}

\bibitem{PDG}
Particle Data Group, C. Caso et al., Eur. Phys. J. {\bf C3} (1998) 1.

\bibitem{secondordersigma}
R. Hamberg, W.L. van Neerven, T. Matsuura, Nucl. Phys. {\bf B359}
(1991) 343;\\
W.L. van Neerven and E.B. Zijlstra, Nucl. Phys. {\bf B382} (1992) 11. 

\bibitem{Wplusfourjets}
F.A. Berends, H. Kuijf, B. Tausk and W.T. Giele,
Nucl. Phys. {\bf B357} (1991) 32. 

\bibitem{resummed}
Yu.L. Dokshitzer, D.I. Dyakonov and S.I. Troyan,
Phys. Rep. {\bf 58} (1980) 269;\\
F. Halzen, A.D. Martin, D.M. Scott and M.P. Tuite,
Z. Phys. {\bf C14} (1982) 351;\\
G. Altarelli, R.K. Ellis, M. Greco and G. Martinelli,
Nucl. Phys. {\bf B246} (1984) 12;\\
J.C. Collins, D.E. Soper and G. Sterman, Nucl. Phys. {\bf B250} 
(1985) 199;\\
C.T.H. Davies, B.R. Webber and W.J. Stirling, 
Nucl. Phys. {\bf B256} (1985) 413;\\
P.B. Arnold and R. Kauffman, Nucl. Phys. {\bf B349} (1991) 381;\\
G.A. Ladinsky and C.-P. Yuan, Phys. Rev. {\bf D50} (1994) 4239;\\
R.K. Ellis and S. Veseli, Nucl. Phys. {\bf B511} (1998) 649. 

\bibitem{AGIS} 
B.~Andersson, G.~Gustafson, G.~Ingelman and T.~Sj\"ostrand, 
Phys.~Rep.\ {\bf 97} (1983) 31.

\bibitem{matching}
H. Baer and M.H. Reno, Phys. Rev. {\bf D44} (1991) 3375;\\ 
J. Andr\'e and T. Sj\"ostrand, Phys. Rev. {\bf D57} (1998) 5767;\\
F.E. Paige, S.D. Protopopescu, H. Baer and X. Tata, hep-ph/9810440.

\bibitem{mike}
M.H. Seymour, Computer Phys. Commun. {\bf 90} (1991) 95.

\bibitem{herwig}
G. Marchesini, B.R. Webber, G. Abbiendi, I.G. Knowles, M.H. Seymour and 
L. Stanco, Computer Phys. Commun. {\bf 67} (1992) 465.

\bibitem{merging}
M. Bengtsson and T. Sj\"ostrand, Phys. Lett. {\bf B185} (1987) 435;\\
G. Gustafson and U. Pettersson, Nucl. Phys. {\bf B306} (1988) 746.

\bibitem{miu}
G. Miu, LU TP 98--9, hep-ph/9804317.

\bibitem{isrshower} 
T. Sj\"ostrand, Phys. Lett. {\bf 157B} (1985) 321;\\
M. Bengtsson, T. Sj\"{o}strand and M. van Zijl, 
Z. Phys. {\bf C32} (1986) 67. 

\bibitem{pythia} 
T. Sj\"{o}strand, Computer Phys. Commun. {\bf 82} (1994) 74.

\bibitem{DGLAP}
V.N. Gribov and L.N. Lipatov, Sov. J. Nucl. Phys. {\bf 15} (1972) 438;\\
G. Altarelli and G. Parisi, Nucl. Phys. {\bf B126} (1977) 298;\\
Yu.L. Dokshitzer, Sov. Phys. JETP {\bf 46} (1977) 641.

\bibitem{Wsigma}
H. Fritzsch and P. Minkowski, Phys. Lett. {\bf 73B} (1978) 80;\\
G. Altarelli, G. Parisi and R. Petronzio, Phys. Lett. {\bf 76B} (1978) 
351;\\
K. Kajantie and R. Raitio, Nucl. Phys. {\bf B139} (1978) 72;\\
F. Halzen and D.M. Scott, Phys. Rev. {\bf D18} (1978) 3378.

\bibitem{D0data} 
CDF Collaboration,F. Abe et al., Phys. Rev. Lett. {\bf 66} (1991) 2951;\\ 
D0 Collaboration, B. Abbott et al., Phys. Rev. Lett. {\bf 80} (1998) 5498.

\bibitem{primkt}
S. Frixione, M.L. Mangano, P. Nason and G. Ridolfi,
Nucl. Phys. {\bf B431} (1994) 453;\\
L. Apanasevich et al., hep-ph/9808467.

\bibitem{Wjet}
D0 Collaboration, B. Abbott et al., FERMILAB-Conf-97/369-E;\\
CDF Collaboration, F. Abe et al., Phys. Rev. Lett. {\bf 81} (1998) 1367.

\end{thebibliography}
\end{document}